\documentclass[useAMS,usenatbib]{mn2e}
\newcommand{\Msun}{\mbox{$M_{\odot}$}}
\newcommand{\Rsun}{\mbox{$R_{\odot}$}}
\newcommand{\Mjup}{\mbox{$M_{\rm{Jup}}$}}

\newcommand{\alphaturb}{\mbox{$\alpha_{\rm{turb}}$}}
\newcommand{\amax}{\mbox{$a_{\rm{max}}$}}
\usepackage{graphicx}
\usepackage{natbib}
\usepackage{amssymb}
\usepackage{hyperref}
\usepackage{amsmath}
\usepackage{lscape}
\usepackage{longtable}
\usepackage{caption}
\usepackage{color}
\usepackage{aas_macros}
\hypersetup{colorlinks=true,citecolor=blue}
\title[Constraining \alphaturb\, in transitional discs with planets. ]{Constraining turbulence mixing strength in transitional discs with planets using SPHERE and ALMA}
\author[M. de Juan Ovelar et al.]{M. de Juan Ovelar$^1$\thanks{m.dejuanovelar@ljmu.ac.uk}, P. Pinilla$^{2}$, M. Min$^{3,4}$, C. Dominik$^{4}$ and T. Birnstiel$^{5}$\\
$^{1}$ Astrophysics Research Institute, Liverpool John Moores University, 146 Brownlow Hill, Liverpool L3 5RF, UK\\
$^{2}$ Leiden Observatory, Leiden University, P.O. Box 9513, 2300RA Leiden, The Netherlands \\
$^{3}$ SRON Netherlands Institute for Space Research, Sorbonnelaan 2, 3584 CA Utrecht, The Netherlands\\
$^{4}$ Anton Pannekoek Institute for Astronomy, University of Amsterdam, 1090 GE Amsterdam, The Netherlands\\
$^{5}$ Harvard-Smithsonian Center for Astrophysics, 60 Garden Street, Cambridge, MA 02138, USA
}
\begin{document}

\date{Accepted 2016 March 21. Received 2016 March; in original form 2015 September 16}

\pagerange{\pageref{firstpage}--\pageref{lastpage}} \pubyear{2014}

\maketitle

\label{firstpage}

\begin{abstract}
We investigate the effect that the turbulent mixing strength parameter \alphaturb\:plays on near-infrared polarimetric and sub-millimetre interferometric imaging observations of transitional discs with a gap carved by a planet.  We generate synthetic observations of these objects with ALMA and VLT/SPHERE-ZIMPOL by combining hydrodynamical, dust evolution, radiative transfer and instrument models for values of  {$\alphaturb=[10^{-4}, 10^{-3}, 10^{-2}]$}. We find that, through a combination of effects on the viscosity of the gas, the turbulent mixing and dust evolution processes, \alphaturb\  strongly affects the morphology of the dust distribution that can be traced with these observations.
We constrain the value of \alphaturb\: to be within an order of magnitude of $10^{-3}$ in TD sources that show cavities in sub-mm continuum images while featuring continuous distribution of dust or smaller cavities in NIR-polarimetric images.

\end{abstract}


\section{Introduction}
\label{sec:intro}
The field of transitional discs (TDs) has recently experienced a paramount push thanks to the technical advancements in high contrast imagers and interferometers. Originally detected and characterised thorough spectral energy distribution (SED) fitting, these protoplanetary discs (PPDs) appeared to be depleted of material in the inner regions and are considered a transition between a full protoplanetary disc and a planetary system \citep{strom89}. {This transitional stage may be caused by processes of planet-disc interaction} {\citep[e.g.][]{rice03, papaloizou07}}, or disc evolution {\citep[e.g.][]{dullemond05,alexander07}}.  

{With resolutions of a few AUs at $140\,\rm{pc}$, current facilities such as the Atacama Large Millimeter/sub-millimeter Array (ALMA) or the new planet imager {Spectro-Polarimetric High-contrast Exoplanet Research} {SPHERE) on the Very Large Telescope (VLT) are providing the community with a plethora of images of TDs showing very different and complex structures {such as rings, azimuthal asymmetries, dips and spiral arms} \citep[e.g.~][]{quanz12,garufi13,vandermarel13, casassus13, perez14,zhang14,walsh14,benisty15,canovas16}. This has triggered a large number of theoretical {studies to explore the potential mechanisms responsible{ {\citep[e.g.][]{regaly12, ataiee13, birnstiel13, zhu14, juhasz15, flock15}.} }}}

{In particular, a group of TDs seem to feature gaps in $870\mu \rm{m}$ interferometric images \citep{andrews11} while showing smaller or non-existent gaps in H-band ($1.2\,\mu\rm{m}$) polarimetric images (i.e.~the \textit{``missing cavities"} problem \citealt{dong12}). With the former tracing the emission of relatively large ($\sim1\,\rm{mm}$) dust grains, and the latter tracing scattering of light by small ($\sim1\,\mu\rm{m}$) ones, these observations suggest that a filtration mechanism is causing the localised depletion of large dust grains, leaving small ones unaffected. Theoretical studies such as \citet{zhu12} or \citet*{pinilla12b}, show that this preferential filtration of certain sizes of dust grains can be caused by a planetary-mass companion while it remains difficult to explain by disc evolution processes. Based on the models presented in the latter, \citet{dejuanovelar13} added radiative transfer and instrument modelling to produce synthetic observations of this scenario showing that images at NIR and sub-mm wavelengths would indeed show this apparent dichotomy and that their combination can be used to estimate the mass of the companion. However, parameters such as the turbulence mixing strength (\alphaturb) are known to have an important effect on the (hydro)dynamical, dust evolution, and radiative transfer processes that govern the evolution of PPDs, and their response to external perturbations \citep[e.g.][]{lyndenbell74,dejuanovelar12,rosotti14}, but its effect on such observations remains to be investigated.} 

{In this letter we explore this issue with general TD models instead of using particular sources. We focus on how such observations can be used to constrain the value of \alphaturb within the range of $10^{-2}$ to $10^{-4}$, currently assumed in the literature and supported by recent observational studies \citep[e.g.][]{mulders12,flaherty15}.} 

The letter is organised as follows:  In \S \ref{sec:models} we outline the modelling procedure.{ In \S \ref{sec:results} we discuss the dust density distribution and synthetic observations obtained, and, in \S \ref{sec:conclusions}, we list the main conclusions of our study.

%%%%%%%%%%%%%%%%%%%%%%%%%%%%%%%%%%%%%%%%%%%%
\section{Models}
\label{sec:models}

\begin{figure*}
\centerline{\includegraphics[scale=0.46]{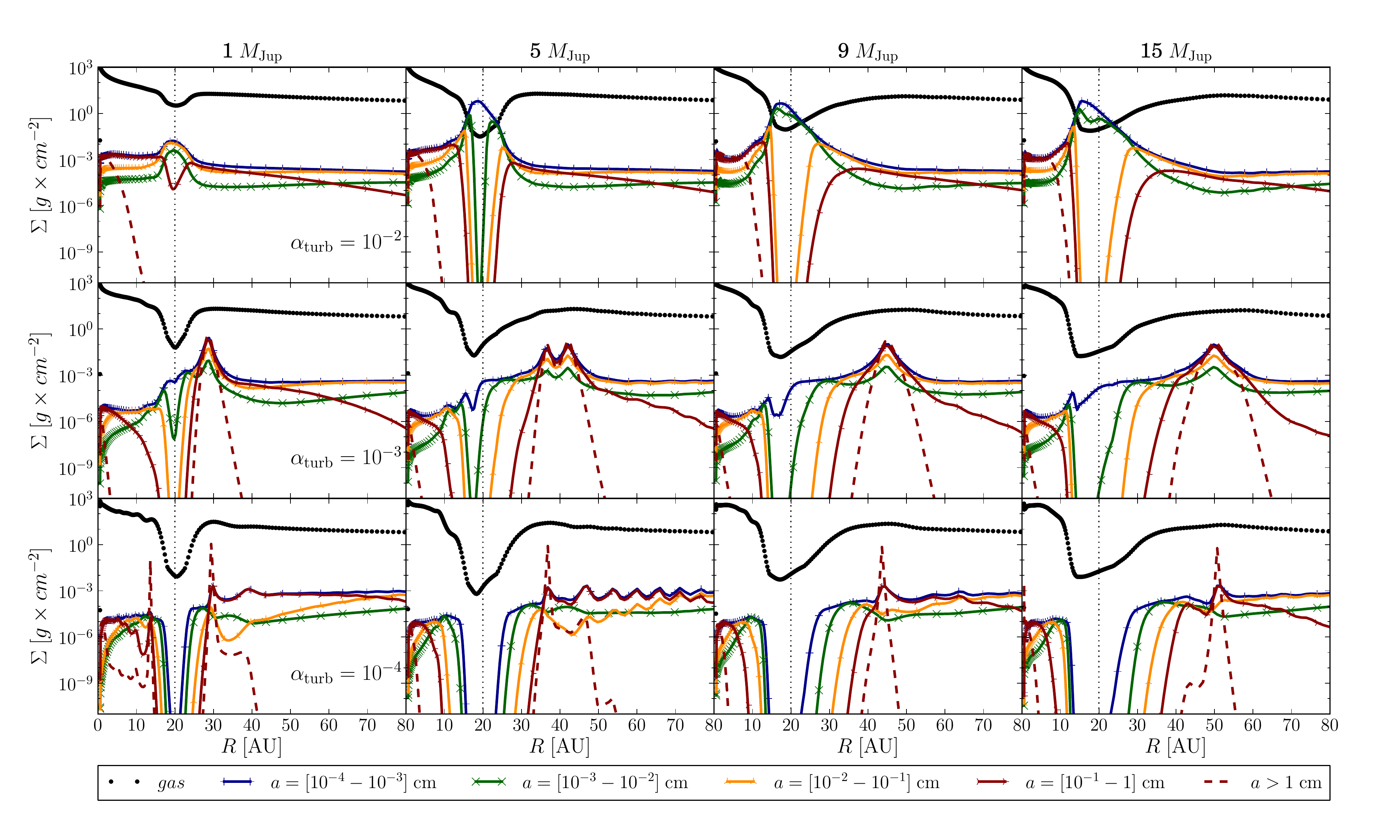}}
\caption{Radial profile of the gas (black diamonds) and dust (coloured lines) distributions for {$\alphaturb=10^{-4}, 10^{-3}$, and  $10^{-2}$ (from bottom to top)}. The blue-plus, green-cross, yellow-vertical-triangle, red-horizontal-triangle and red-dashed lines correspond to dust bin sizes of $a=[10^{-4}-10^{-3}]\,\rm{cm}$, $a=[10^{-3}-10^{-2}]\,\rm{cm}$, $a=[10^{-2}-10^{-1}]\,\rm{cm}$, $a=[10^{-1}-1]\,\rm{cm}$, and $a>1\,\rm{cm}$, respectively. {Note that the piling up of small grains at the planet's location in \alphaturb$=10^{-2}$ cases is a numerical artifact (see details in Section \ref{sec:dustexplained})}}
\label{fig:dustbins}
\end{figure*}

Following the same methodology {and models presented in \citet{dejuanovelar13}}, we combine {2D-hydrodynamical, 1D-dust evolution}, radiative transfer, and instrument simulations to produce synthetic observations of a disc hosting a planet of masses $M_{\rm{p}}=[1,5, 9,15]\,\rm{M_{jup}}$. The values of all parameters are given in Appendix \ref{sec:apmodels} together with a brief description of the modelling procedure. For more details of our method we refer the reader to the above mentioned paper. We run all cases with three values of $\alphaturb=[10^{-4}, 10^{-3}, 10^{-2}]$.

\section{Results}
 \label{sec:results}
 
{In the interest of space} we describe the effect of \alphaturb\: on the gas and dust distribution in all cases while discussion on images is focussed on models run with $1$ and $9\,\Mjup$ planets only, which are representative of our sample. Synthetic images of cases $[5,15]\,\Mjup$ are shown in Appendix \ref{sec:appimages}.

\begin{figure*}
   \centering
   \begin{tabular}{c} 
   \includegraphics[width= 75mm,height=110mm]{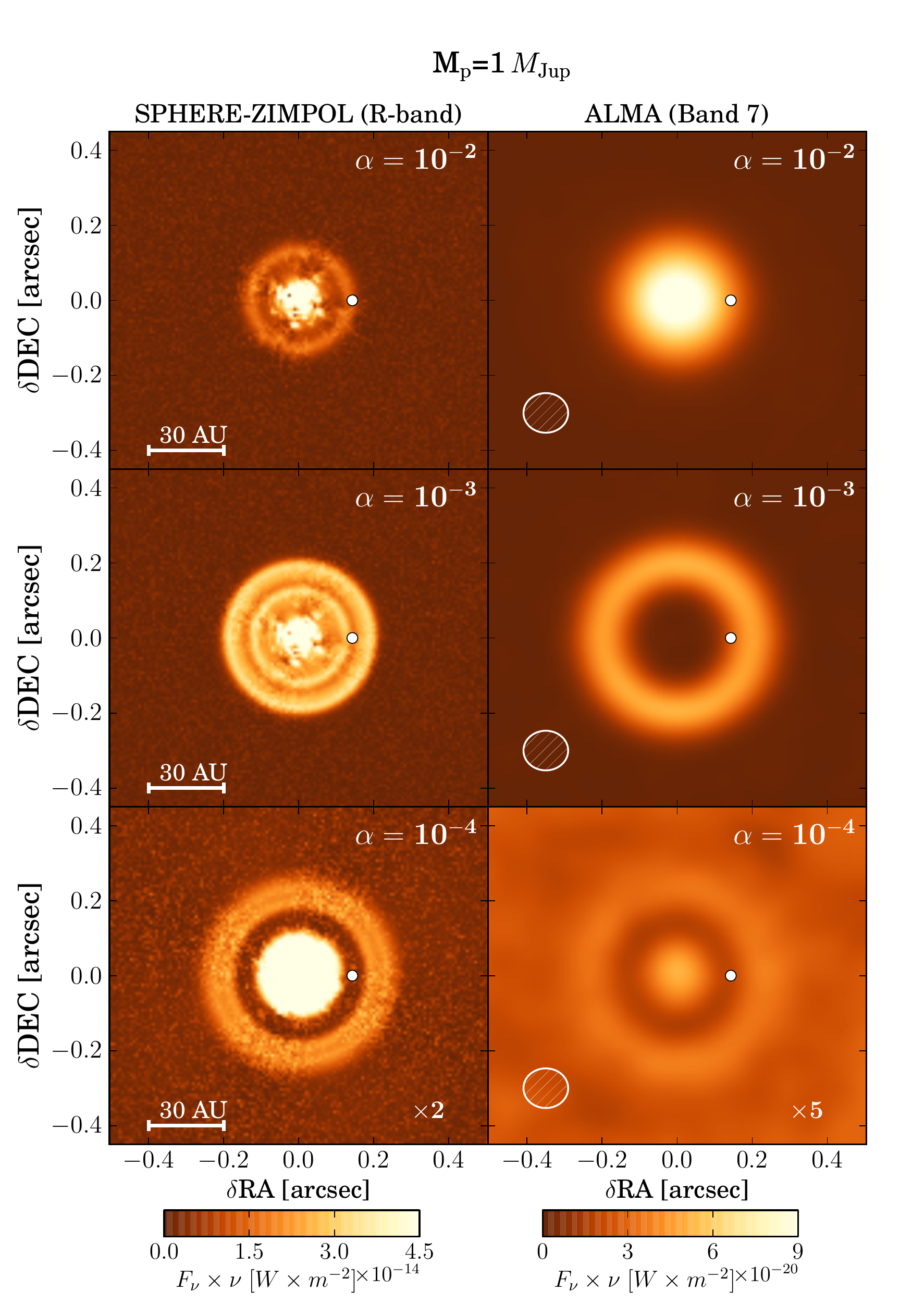}\hspace{0.8mm} 
   \includegraphics[width= 75mm,height=110mm]{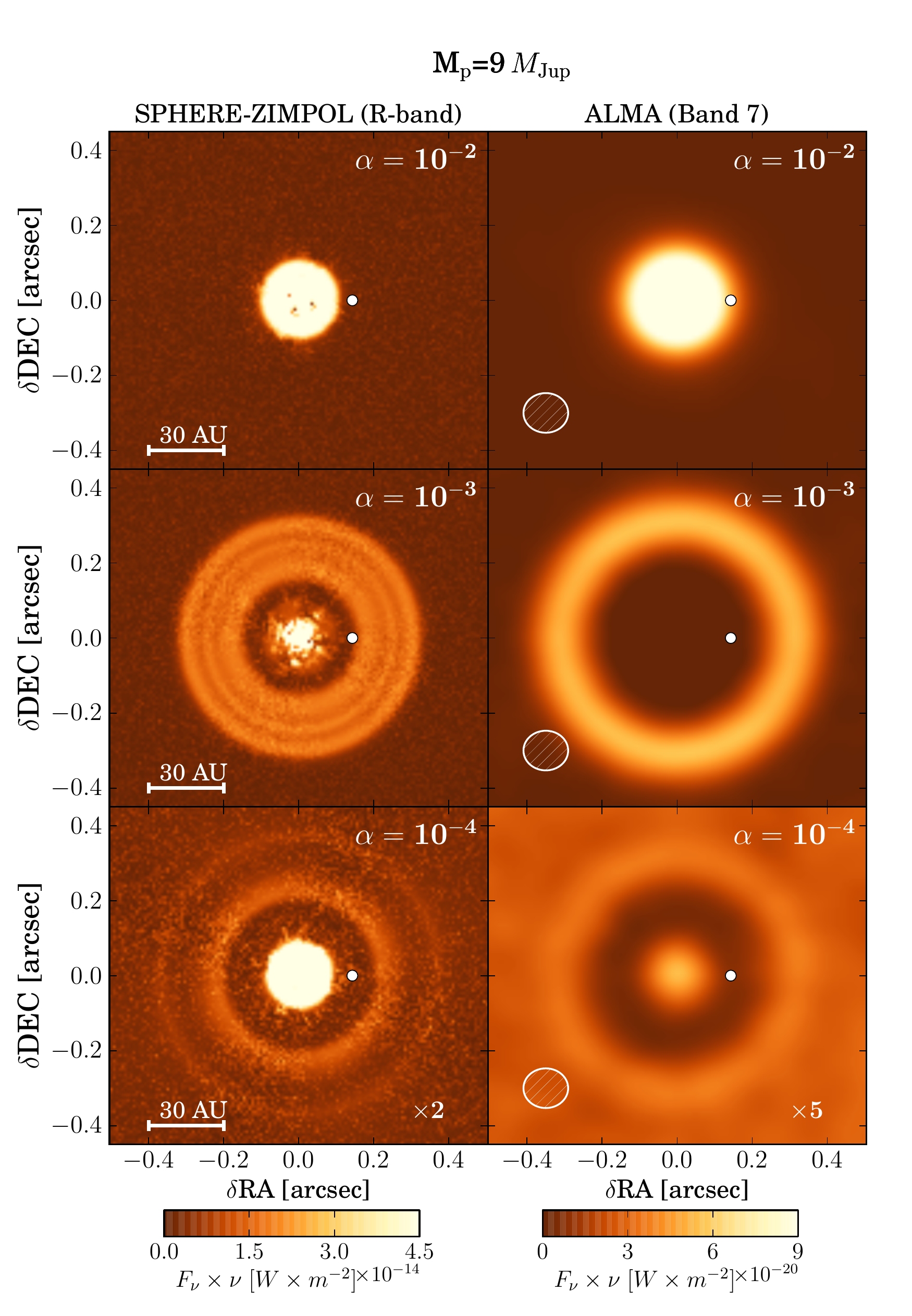}   
    \end{tabular}	
   \caption{R-band SPHERE and ALMA Band 7 synthetic observations of a protoplanetary disc with a $1$ and $9\Mjup$ planet embedded at $20\,\rm{AU}$ for the three values of \alphaturb\:considered. The flux in the case of $\alpha_{\rm{turb}}=10^{-4}$ has been increased by of 2 and 5 for the SPHERE and ALMA images, respectively. {Note that the ring-like feature in the case of $1\Mjup$ and \alphaturb$=10^{-2}$ is a numerical artifact (see details in Section \ref{sec:dustexplained}).} }
   \label{fig:images}
\end{figure*}

\subsection{Impact of \alphaturb\,\,on the gas and dust density distribution}
\label{sec:dustexplained}

Figure \ref{fig:dustbins} shows the radial profiles of the gas and (binned by size from $1\,\rm{\mu m}$ to $>1\,\rm{cm}$) dust distributions in the disc for the different simulations. Panels in {each column show the three cases of \alphaturb\:studied for each planetary mass considered}. {The gap-opening power of a planet is determined by the balance between the mutually counteracting gravitational and viscous torques that arise from its presence in the disc {and the viscous conditions of the gas, respectively}\citep[e.g.][]{crida06}. Thus, as the value of \alphaturb\:decreases (i.e.~upper to lower panels), the gap opened by a planet of a certain mass in the gas distribution is significantly deeper. The gap also becomes wider as planet mass increases (for a fixed \alphaturb). Additionally, the pressure gradient becomes positive at the outer edge of the gap opened{ and a pressure maximum appears} {\citep[e.g.][]{paardekoper04}}. {The characteristics of the new pressure gradient distribution (e.g.~steepness) and those of the pressure maximum (e.g.~amplitude) control the filtering/trapping of dust particles of different size in the disc} \cite[see Equation 11 in][]{pinilla12b}.}}

{Turbulent mixing also plays an essential role on dust growth and evolution. For instance, fragmentation occurs because of high relative velocity collisions between dust grains, with main sources being turbulent motion and radial drift. If radial drift is reduced by a positive pressure gradient, turbulent motion dominates and the maximum grain size that particles can reach before fragmentation (\amax) depends directly on \alphaturb\:\citep{birnstiel10a,birnstiel12}. If  \alphaturb\:is high, \amax\: can be much lower than the size of particles that can be trapped in pressure bump, preventing accumulation of mm-sized particles in the pressure bump \citep[e.g.][]{pinilla15}. In addition, turbulence also drives diffusion of particles within pressure bumps, and therefore if  \alphaturb\:is high, particles can more easily \textit{escape} a dust trap. This is the case of models with $\alphaturb=10^{-2}$ (upper row) where we see no dust traps for mm- or even cm-sized particles, even for very massive planets.} In summary, there are three reasons why trapping is weaker for higher values of  \alphaturb: less deep planetary gaps and hence lower pressure gradient at the outer gap edge; more effective fragmentation of particles which leads to smaller grains that are more difficult to trap; and higher diffusion or mixing of dust that allows the particles to escape from the trap. 

{{Additionally, when \alphaturb$=10^{-2}$, and also independently from planet mass, small grains pile up at the position of the planet even surpassing the surface density values of the gas. This is an effect of our 2D (gas)+1D (dust) approximation, in particular from assuming the gas velocity using viscous accretion and assuming the averaged gas surface density for the dust evolution as in \citet{pinilla12b,pinilla15}. To test this, we re-run the dust calculations in the $1\Mjup$ and $\alphaturb=10^{-2}$ case including only dynamics and neglecting coagulation, fragmentation and grain growth processes. The enhancement of small particles then remains suggesting that it is indeed a numerical artifact and not the result of dust evolution processes. We then run another simulation for this case, where the dust velocities are assumed to be $\overline{v_{\rm{gas}}* \Sigma_{\rm{gas}}} / \overline{\Sigma_{\rm{gas}}}$ instead, with $v_{\rm{gas}}$ and $\Sigma_{\rm{gas}}$ being the azimuthally and time (over the last 100 orbits) averaged values from the hydrodynamical simulations. In this case, accretion rate throughout the gap is almost constant and the pile up of small dust at the location of the planet disappears (see details in Appendix \ref{sec:appimages}). Because in our dust evolution models we assume that the gas velocity comes from viscous evolution that tries to close the gap, the particles that feel these velocities are pushed into the gap. Since the viscous velocities are proportional to \alphaturb, in the case of lower values (i.e.~$10^{-3}$, $10^{-4}$) they are negligibly small, and radial drift becomes the dominating contribution for the dust velocity, which moves dust up the pressure gradient and prevents this artifact from appearing. {This is, therefore, an inherent limitation of our modelling procedure that affects high turbulence cases in the region of the disc near the planet. To treat this issue, 2D gas and dust evolution models that include grain growth and dynamics simultaneously are needed, which are beyond the scope of our study.
{For our analysis of these cases we therefore ignore this feature and base our conclusions on the otherwise continuously decreasing distribution of dust.}} 

We also note that our dust evolution treatment cannot follow processes occurring when dust-to-gas ratios are larger than 1 which can trigger instabilities and fast growth to planetesimals \citep[e.g.][]{johansen07}.}}

{In the cases where \alphaturb\:is low ({$\alphaturb=10^{-4}$, lower panels in Figure \ref{fig:dustbins}}), the effect on the gas distribution would in principle favour trapping. However, dust growth is increased because turbulent relative velocities are low: growth dominates over fragmentation and particles can continue growing to even meter-sized objects inside the trap. Additionally, turbulent diffusion is not effective. As a result, only particles larger than cm-sizes accumulate in a very narrow ring {(red-dashed line in Fig.~\ref{fig:dustbins})} and lower dust grain sizes are depleted. The {waves that appear with low values of  \alphaturb\ (e.g.~$> 40\,$AU features in $5\,\Mjup, \alphaturb=10^{-4}$ case) are an artifact. {These features come from the spiral waves in the hydrodynamical simulations that appear as fixed density bumps to the dust evolution code because we assume a stationary gas distribution azimuthally averaged after 1000 orbits. However, they have a pattern speed equal to the planet, and thus are unable to trap dust}.}  

All results are compared after 1~Myr of dust evolution. 
 
\subsection{SPHERE-ZIMPOL/ALMA images for different values of \alphaturb}
\label{sec:images}

The combination of effects of \alphaturb\:on the gas and dust distribution of the disc has a clear impact on scattering and emission flux images. Figure \ref{fig:images} shows synthetic SPHERE-ZIMPOL R-band ($0.65\mu \rm{m}$) polarimetric and ALMA Band 7 ($850\mu \rm{m}$) continuum emission observations (first and second columns of each panel, respectively) of a disc with an embedded $1$ (left panel) and $9\Mjup$ (right panel) planet and for \alphaturb $=[10^{-4},10^{-3},10^{-2}]$ (bottom to top rows). 

\subsubsection{Case of $\alphaturb=10^{-2}$}

{{In the $1\Mjup$ and \alphaturb$=10^{-2}$ case (top row) there is no effective trapping and dust grains of all sizes populate the region with approximately constant surface density up to the location of the planet where small particles are artificially enhanced due to {the limitations of our model} (see previous subsection). {}Unfortunately, our simulations of polarimetric observations of SPHERE-ZIMPOL at short wavelengths ($\sim0.65\,\mu\rm{m}$) are dominated by this feature in this particular case. These images trace starlight scattered by small ($\sim1\,\mu\rm{m}$, blue and green lines in Fig.~\ref{fig:dustbins}) dust grains at the surface of the disc, and therefore show the abrupt enhancement in density of the small grains in this location as a narrow ring. When ignoring the pile-up, the distribution of small ($<1\,\rm{mm}$) grains remains {rather constant, and, therefore, one would expect this to show in the NIR image as a continuous disc. To clarify this, we show in Appendix \ref{sec:appimages} SPHERE and ALMA images of this case when using a manually smoothed prescription for the velocities inside the gap. All simulations of $M_{\rm{p}}>1\Mjup$ and this value of \alphaturb\: show the inner region of the disc because the gradient of the small grains distribution here is very high. It is this gradient instead of the pile-up feature what dominates the image in these cases.}}}

{{ALMA images trace thermal emission from $\sim1\,\rm{mm}$ grains (yellow and red lines in Fig.~\ref{fig:dustbins}) and therefore show a continuous distribution with large grains more depleted in the outer regions of the disc. As the planet becomes more massive ($9\Mjup$ panel in Fig.\ref{fig:images}) the gap becomes deeper and wider but {the pressure maximum at the outer edge is} still not strong enough to trap efficiently {and, therefore, no ring-like feature appears in these images}.}}

\subsubsection{Case of $\alphaturb=10^{-3}$}

For an intermediate value of \alphaturb$=10^{-3}$ (middle rows in Fig.~\ref{fig:dustbins} and \ref{fig:images}) the situation changes. Trapping here is effective for dust grains of sizes around $a\sim1\,\rm{mm}$ and therefore, ALMA continuum images show a ring of dust where the dust trap is located (i.e.~at the location of the pressure bump). {On the other hand, {for a $1\Mjup$, } small particles -- still coupled to the gas -- flow freely through the gap to the inner regions of the disc ($r<r_{\rm{planet}}$).} SPHERE-ZIMPOL images therefore show two components for the disc separated by the gap at the location of the planet, where small grains are partially depleted and cannot scatter as much radiation. For a $9\Mjup$ planet the effects are amplified in both images. The trapping power of the planet is much larger, the pressure bump moves outwards and the inner regions to the position of the planet are strongly depleted. This causes SPHERE-ZIMPOL images to show very clearly the position of the wall (outer edge of the gap) in the disc whose surface is covered with small grains well coupled to the gas and effectively scattering starlight. The pressure bump trapping large grains shines in ALMA continuum emission images as a wide ring at around $\sim45\,\rm{AU}$.

\subsubsection{Case of $\alphaturb=10^{-4}$}

The models with the lowest value \alphaturb\:we consider ($\alphaturb=10^{-4}$) tell the story of dust coagulation. Despite the fact that turbulence here is very low and therefore the effect of the planet on the gas distribution is amplified (which favours effective trapping), relative velocities between dust grains are very low and coagulation processes dominate over fragmentation ones in the dust trap. This results  in a distribution of grain sizes dominated by larger than cm sizes (red-dashed line in Fig.~\ref{fig:dustbins}) which leads to low fluxes at sub-mm wavelengths. The dust trap affecting mm grains is present in images of ALMA in both planet mass cases, but it is an extremely faint feature. The gap opened by a $1\Mjup$ planet appears in the SPHERE-ZIMPOL image thanks to the fact that it traces scattering radiation instead of emission, which will be affected by the strong depletion of small grains due to coagulation. Indeed, few grains still scatter efficiently from the surface of the disc and wall, and are therefore able to trace the gap. Note that in NIR and sub-mm images obtained for this value of \alphaturb\: the flux has been increased by factors of $[2,5]$, respectively. When a more massive planet of $9\Mjup$ is opening the gap however, the depletion becomes very strong and the image, although still tracing the wall, becomes much fainter. Here the NIR-scattering image shows a secondary ring corresponding to one of the artifacts mentioned in the previous section.

\section{Conclusions}
\label{sec:conclusions}

{We perform (2D-)hydrodynamical, (1D-)dust evolution}, radiative transfer and instrument simulations to obtain synthetic SPHERE-ZIMPOL and ALMA observations of TDs where a gap is opened by a planet of different masses and with three different values of the turbulent mixing strength parameter $\alphaturb=[10^{-4},10^{-3},10^{-2}]$.  {In this work we do not have simultaneous evolution of gas and dust, but assume the gas density from hydrodynamical simulations of planet disc interaction after 1000 orbits. The gas density profile remains then static for the dust evolution and we do not include planet accretion or migration (see Appendix \ref{sec:apmodels} for details). Under these assumptions, we find} that \alphaturb\: has a major impact on observations of dust in the disc. In particular our results show that:}

\begin{itemize}

\item{{{We confirm that, as shown in \citet{pinilla12b,pinilla15}, for $\alphaturb=10^{-2}$ the trapping mechanism is weak, resulting in SPHERE-ZIMPOL and ALMA images showing continuous distributions of $\sim1\,\mu\rm{m}$ and $\sim$mm dust grains, respectively {(see text for details on the limitations of our models and the $1\Mjup$ image in this case)}.}}} 

\item{For values of \alphaturb$=10^{-4}$, growth is favoured over fragmentation, and dust grains grow to sizes of $\gtrsim1\,\rm{cm}$ inside pressure traps, resulting in very faint fluxes and a a gap/ring-like structure in both sub-mm and NIR polarimetric images.}

\item{Current observations of TDs showing continuous distributions (or small gaps/cavities) in NIR-polarimetric images, and large gaps/cavities and ring-like features in sub-mm images (i.e.~the \textit{"missing cavities''} effect) are only reproduced when \alphaturb$=10^{-3}$. {Since, to our knowledge, no mechanism other than planet-disc interaction has been proposed to cause this effect, it is reasonable to assume that such combination of images is indicative of the presence of a planet. Then, according to our results, the value of \alphaturb\: can be constrained to $10^{-3}$ within an order of magnitude, and the mass estimator presented in \citet{dejuanovelar13} can be used to estimate the mass of the companion in these sources.} {Note that ALMA images on their own could also be used to constrain the value of \alphaturb\: but this is \textit{only if} one knows for sure that a planet is causing the gap and then SPHERE-ZIMPOL images would still be needed to use the planet-mass estimator. } }

\end{itemize}

\section{Acknowledgements}

The authors are thankful to the anonymous referee for a thorough review, and to J.\ M.\ D.\ Kruijssen and G.\ P.\ Rosotti for useful comments on the manuscript. T.\ B. is supported by the NASA Origins of Solar Systems grant NNX12AJ04G and the Smithsonian Institution Pell Grant program. P.\ P. is supported by Koninklijke Nederlandse Akademie van Wetenschappen (KNAW) professor prize to Ewine van Dishoeck.

\bibliographystyle{mn2e}
\bibliography{/Users/mj/Documents/Bib/mjsbib.bib}

\appendix

\section{Modelling approach}\label{sec:apmodels}

\subsection{Gas and dust models}

\begin{table}
	% Stretch the rows a bit to give more space
	\renewcommand{\arraystretch}{1.3}
	\renewcommand{\captionfont}{\scshape}
	% Changes table font (not caption! see above)
	\small	
	\begin{center}
		\caption{Basic parameters of the simulations}\label{tab:discmodel}
		\begin{tabular}{ll}
			\hline
			Temperature of the star ($T_{\rm{star}}$) 			& $4730\,\rm{K}$\\
			Radius of the star ($R_{\rm{star}}$)				& $1.7\,\rm{\Rsun}$\\
			Mass of the star ($M_{\rm{star}}$)				& $1\,{\Msun}$\\
			Mass of the disc ($M_{\rm{disc}}$)				& $0.0525\,\Msun$\\
			Position of the planet ($R_{\rm{p}}$)				& $20\,\rm{AU}$\\
			Fragmentation velocity ($v_{\rm{f}}$)			& $10\,\rm{m/s}$\\
			Inner disc radius ($R_{\rm{disc,inn}}$)			& $0.025\,R_{\rm{p}}$\\
			Outer disc radius ($R_{\rm{disc,out}}$)			& $7.0\,R_{\rm{p}}$\\
			Solid density of dust particles ($\rho_{\rm{dust}}$)	& $1.2\,\rm{g/cm^3}$\\
			Alpha viscosity ({$\alpha_{\rm{turb}}$})& $[10^{-2}, 10^{-3}, 10^{-4}]$\\
			\hline
		\end{tabular}
	\end{center}
\end{table}%
 
 \begin{figure*}
\centerline{\includegraphics[scale=0.55]{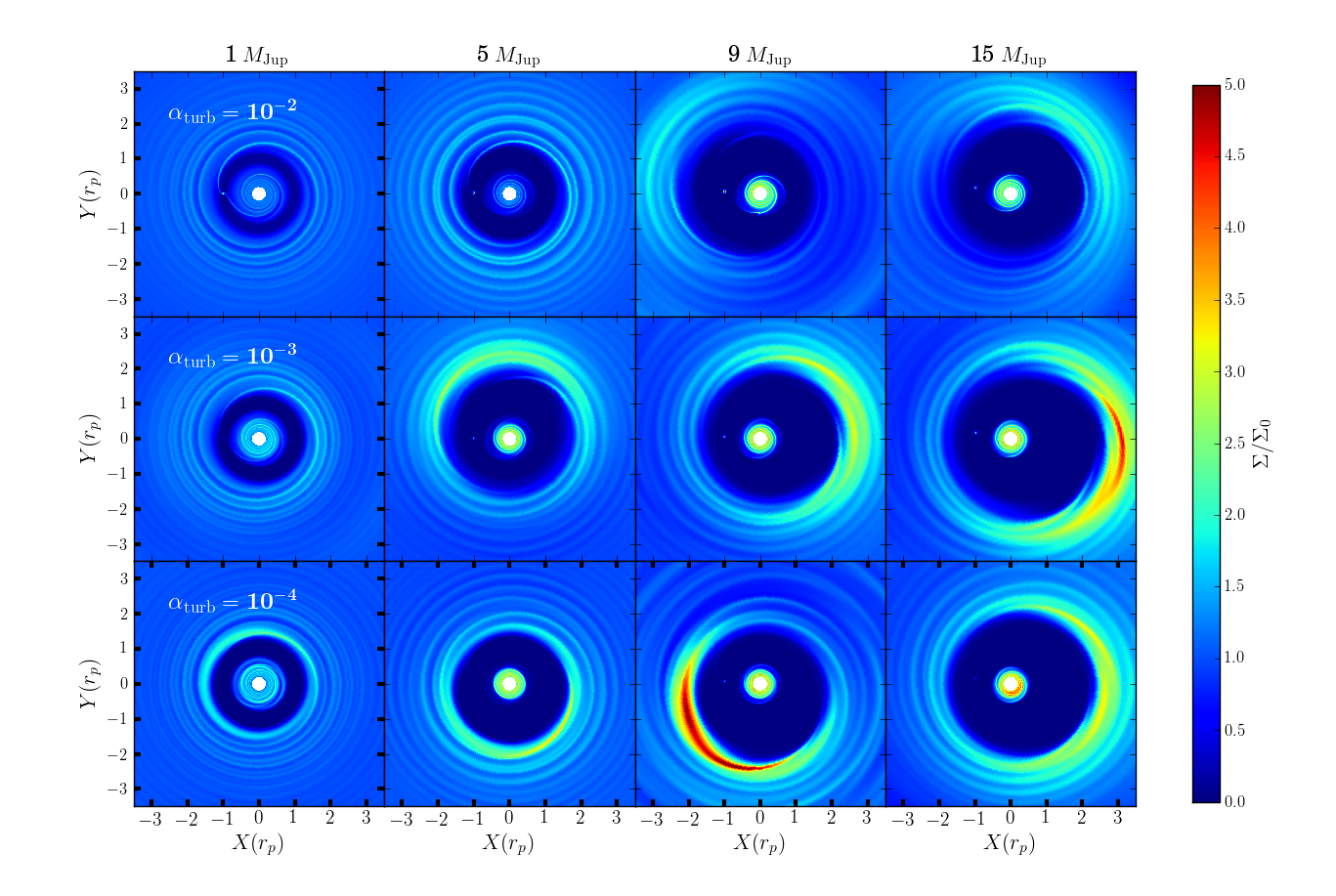}}
\caption{2-D Gas distributions for all cases studied after $1000$ orbits of evolution. {The colorbar shows the surface density distribution of gas in the disc normalised to the initial distribution (i.e.~at orbit zero)}}
\label{fig:gas}
\end{figure*} 
 
 We use the hydrodynamical code FARGO \citep{masset00} to study the evolution of the gas distribution on a 2-D (radial+azimuthal) disc set up with an embedded planet at $R_{\rm{p}}=20\,\rm{AU}$. Table\,\ref{tab:discmodel} shows the parameters used as input to the code (these are the same as in \citealt{dejuanovelar13}). We select open boundary conditions. The logarithmically extended radial grid {has a resolution of 512 in radius by 1024 in azimuth and} is taken from $R_{\rm{disc,inn}}=0.05\,\rm{AU}$ to $R_{\rm{disc,out}}=140\,\rm{AU}$. Here we consider $\Sigma\propto r^{-1}$, a kinematic viscosity of $\nu = \alphaturb c_s^2/\Omega$ \citep{shakura73}, with $c_s$ and $\Omega$ being the sound speed and Keplerian frequency respectably, and  with {$\alphaturb=[10^{-4}, 10^{-3}, 10^{-2}]$}. {The disk is assumed to be a flared disk with $h/r \propto r^{1/4}$ , such that the temperature scales as $T \propto r^{1/2}$. And the aspect ratio at the location of the planet is 0.05.} We set the mass of the disc to $M_{\rm{disc}}=0.0525\,\rm{\Msun}$. Stellar parameters are those typical of a T-Tauri star. 

Figure\,\ref{fig:gas} shows the 2-D surface density distribution of gas in all simulations run after 1000 orbits of evolution. This distribution is azimuthally averaged and fed as input to the 1-D dust evolution code. {Note that this is an important caveat of our method since by doing this we are assuming that the gas distribution remains quasi-static after 1000 orbits, which may not be the case in particular for the cases with very massive planets where strong spiral shocks and/or vortices still exist in the disk after this time.} The shape, depth and width of the gap is clearly dependent on the mass of the planet. {In addition, azimuthal structures are also affected by the viscosity, as for instance the presence of a vortex or eccentric  shape of the outer edge of the planetary gap, which existence depends on the planet mass and disc viscosity \citep{kley06,ataiee13, zhu14, fu14}.}

Density waves are clearly appearing in almost all cases but one should bare in mind that this is just a snapshot and that these waves {cannot act as dust traps.} 

To obtain the dust distribution, as we mentioned earlier, {we take the evolved gas surface density from the hydrodynamical simulations and then compute grain growth and fragmentation in the dust due to radial drift, turbulent mixing, and gas drag forces. We use the 1-D (radial) code described by \citet{birnstiel10a}. The dust is initially distributed {such that the dust mass is $1\%$ that of the gas (after 1000 orbits), with {an} initial size for the grains of $1\,\mu$m and evolving the distribution for $1\,\rm{Myr}$.}  {For the dust, we have a logarithmically spaced grid of the grain size {with 180 cells covering sizes} from $1\,\mu m$ to $200\,$ cm, {and dust densities for each grain size}.}

Note that we do not take into account feedback from the dust onto the gas which becomes non-negligible when the dust-to-gas ratio approaches values close to unity. This effect could cause a secondary ring at mm-wavelengths (result of a secondary pressure bump) even in the case of a disc hosting a single planet, as shown in \citet{gonzalez15}. {Other processes that we do not take into account are planet migration and accretion of dust onto the planet. {Regarding the latter, \citet{owen14a} showed that the luminosity created by the process could affect the SED of the source, but to our knowledge no study has investigated its effect on the general distribution of dust in the disc. In the absence of such study, other than the distribution of dust in the surroundings of the planet and possibly streams connecting the edges of the gap with the planet, we have no reason to believe that the overall morphology of the distribution of dust in the disc  (i.e.~radial location of pressure maxima, gap wall, etc...) will be drastically affected. We do not expect any changes due to migration either, since the timescales for planet migration for such high mass planets will be similar to viscous timescales, meaning that the trap will move as the planet migrate (as explained in \citealt{pinilla15}) which will cause the relative radial location of small and big grains to remain constant.} } {We also performed a test run using typical parameters of a Herbig star instead of a T-Tauri one, finding that the basic morphology of the gas/dust distribution remains the same as in the cases presented here and only the brightness of the images changes.}

\subsection{Radiative transfer and synthetic observations}

We compute full resolution\footnote{(In these images the resolution is not limited by the capabilities of a particular instrument; they are the direct output from the MC Radiative transfer simulation)} emitted and scattered flux images of the dust distribution in the disc at wavelengths of $\lambda=[0.65,850]\,\rm{\mu m}$ to use them as a model input for SPHERE-ZIMPOL and ALMA simulators in bands R and 7, respectively. The radiative transfer is carried out using the Monte-Carlo radiative transfer code MCMax \citep{min09}, {which self-consistently solves the temperature and vertical structure of the distribution of gas and dust in the disc, provided as an input, including the effect of dust settling.}
 
For the opacities of the dust we use a mixture of silicates ($\sim$$58\%$), iron sulphide ($\sim$$18\%$), and carbonaceous dust grains ($\sim$$24\%$) with an average density of $\rho_{\rm{mix}}=3.2\,\rm{g/cm^{3}}$ \citep{min11} and a porosity of $p=0.625$, corresponding to the density $\rho=1.2\,\rm{g/cm^{3}}$ used in the dust evolution simulations. The indices of refraction are obtained from: \citet{dorschner95,henning96} for the silicates, \citet{begemann94} for the iron sulphide, and \citet{preibisch93} for the carbonaceous dust grains. For each value of the viscous turbulence (\alphaturb), MCMax self-consistently solves the vertical structure of the gas in the disc iteratively assuming vertical hydrostatic equilibrium. The vertical structure of the dust is then computed considering settling and turbulent mixing.

We simulate polarimetric and interferometric observations with SPHERE and ALMA at $\lambda=[0.65,850]\,\rm{\mu m}$ with the SPHERE-ZIMPOL \citep{thalmann08} and CASA \citep{CASAref}\footnote{http://casa.nrao.edu/} simulators, respectively. We assume {one hour} of total observing time with both instruments. In the SPHERE-ZIMPOL simulator we select the RI filter and process the full resolution Stokes $Q$, and $U$ images obtaining the final polarised-intensity image as $PI=\sqrt{Q^2+U^2}$. To obtain synthetic observations with ALMA in Band 7 we process the $850\,\rm{\mu m}$ images with the CASA simulator specifying a center frequency of $\nu_{\rm{obs}}=345\,\rm{GHz}$, and a band width of $\Delta \nu_{\rm{obs}}=7.5\,\rm{GHz}$. As for the ALMA configuration used, we specify a desirable resolution of $0,1\arcsec$\footnote{The CASA simulator accepts this as an input and then searches for the available configuration that achieves the specification.} and include the effect of atmospheric noise.

\section{Additional synthetic images} \label{sec:appimages}
\begin{figure*}
   \centering
   \begin{tabular}{c} 
   \includegraphics[width= 80mm,height=120mm]{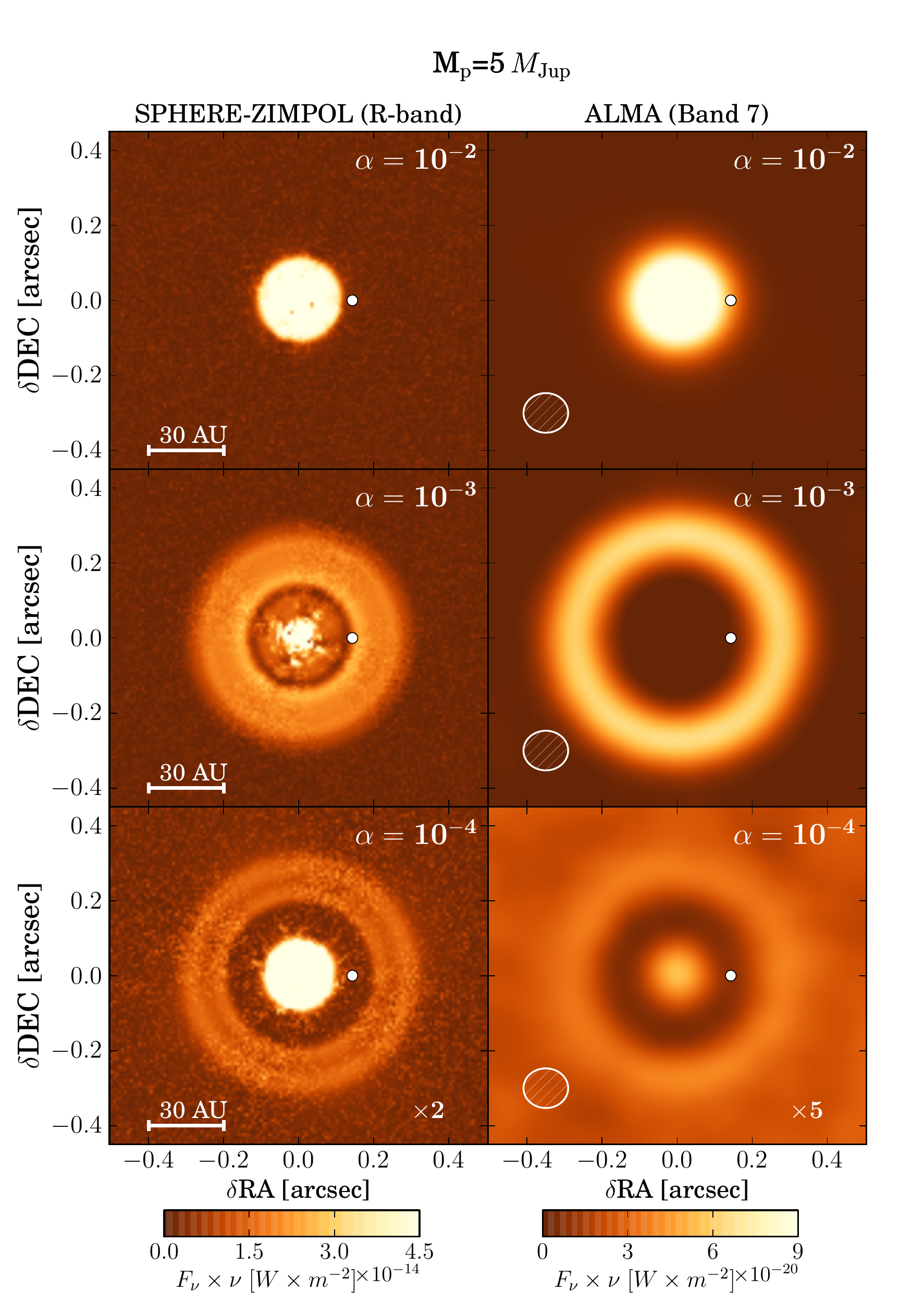}\hspace{0.8mm} 
   \includegraphics[width= 80mm,height=120mm]{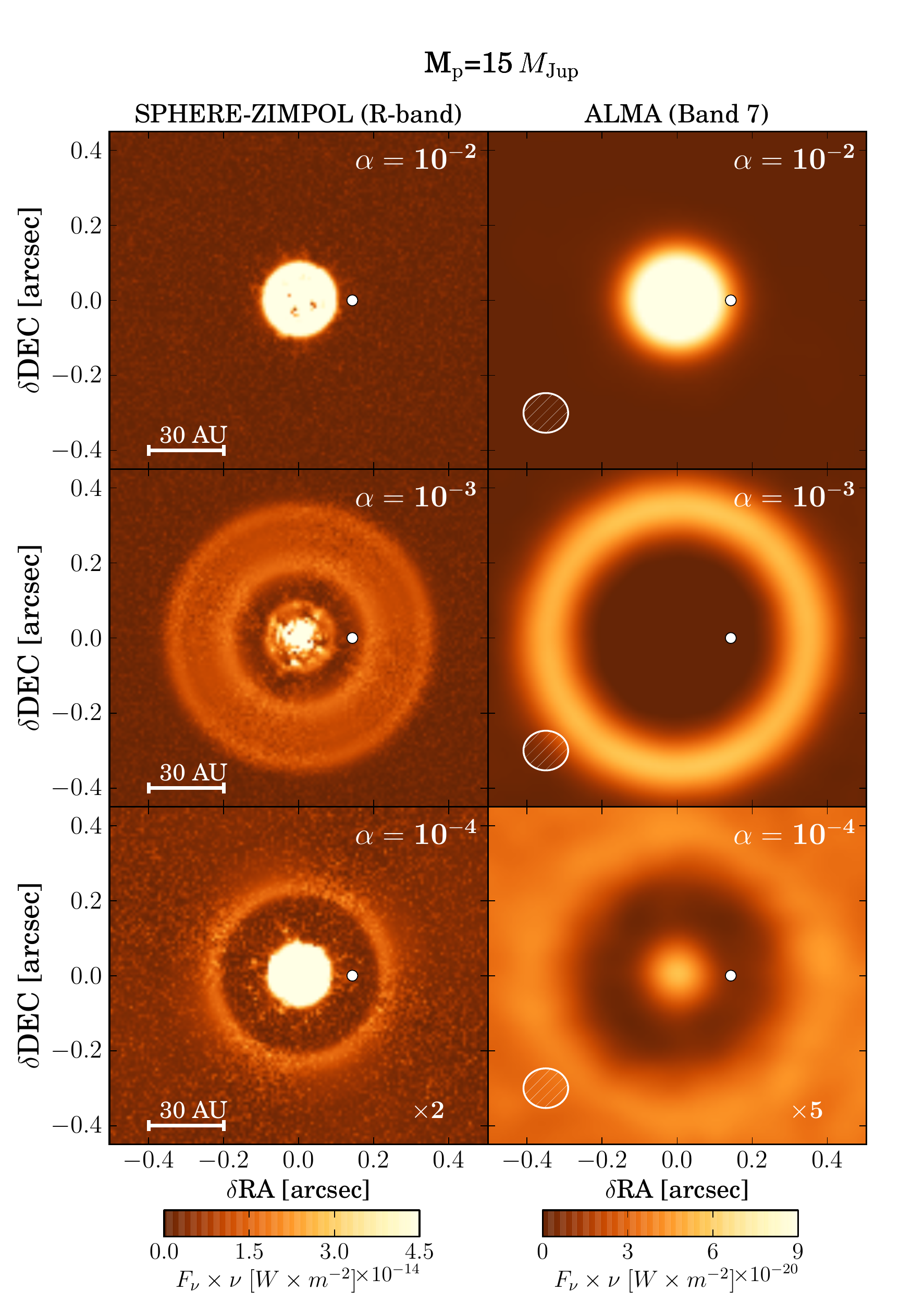}
    \end{tabular}	
   \caption{R-band SPHERE and ALMA Band 7 synthetic observations of a protoplanetary disc with a $5$ and $15\Mjup$ planet embedded at $20\,\rm{AU}$ for the three values of \alphaturb\:considered. The flux in the case of $\alpha_{\rm{turb}}=10^{-4}$ has been increased by of 2 and 5 for the SPHERE and ALMA images, respectively.}
   \label{fig:imagesapp}
\end{figure*}

Figure \ref{fig:imagesapp} shows how the cases of planets of masses $5$ and $15\,\Mjup$ follow the same trends pointed out in Section \ref{sec:results}. Only the cases of \alphaturb$=10^{-3}$ show the missing cavities feature, i.e.~no gap (or small gap) in NIR-scattering images while large gap in sub-mm ones.

{{Figure \ref{fig:imagesTEST} show the synthetic observations for the case of $1\Mjup$ and \alphaturb$=10^{-2}$ when the gas velocity considered for the dust evolution inside the gap is $\overline{v_{\rm{gas}}* \Sigma_{\rm{gas}}} / \overline{\Sigma_{\rm{gas}}}$ instead of the one set by viscous accretion (see Section \ref{sec:dustexplained}). This prescription prevents the small grains from piling-up at the position of the planet as a consequence of the viscous velocities pushing them towards this position. {However, this is a manual adjustment, and a physically reasonable approximation for cases of a low mass companion ($\lesssim1\Mjup$ planet), while it becomes more of a challenge for more massive planets because of the large velocity fluctuations close to the planet. For consistency with our modelling approach, we keep the results from all our models in the main text and only show here the alternative image for clarification purposes.} When the artifact is removed, the observations clearly show the otherwise continuous dust distribution in both SPHERE-ZIMPOL and ALMA observations at NIR and sub-mm wavelengths, respectively.}}

\begin{figure}
   \includegraphics[scale=0.55]{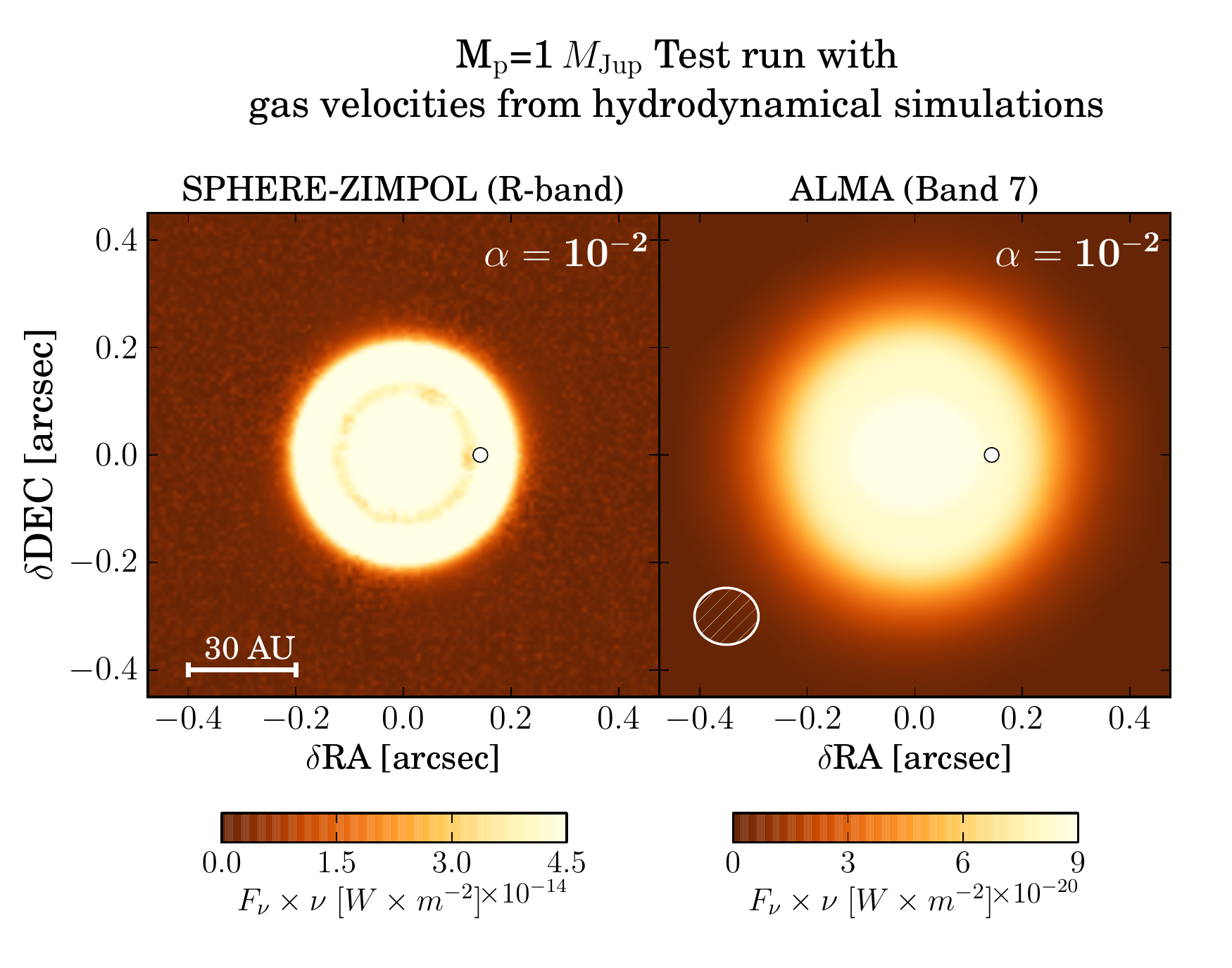}
   \caption{R-band SPHERE and ALMA Band 7 synthetic observations of $1\Mjup$ planet and \alphaturb$=10^{-2}$ case with the $\overline{v_{\rm{gas}}* \Sigma_{\rm{gas}}} / \overline{\Sigma_{\rm{gas}}}$ prescription considered for the gas velocity in the gap instead of the viscous accretion one (see Section \ref{sec:dustexplained}).}
   \label{fig:imagesTEST}
\end{figure}

\end{document}